# Corrections to universal scaling in real maps

Keith Briggs
Department of Mathematics,
University of Melbourne,
Parkville, Australia 3052. [*]
kbriggs@maths.adelaide.edu.au

May 11, 1994

**Abstract**

I discuss the universal aspects of scaling in period-doubling sequences in families of maps of the real line possessing non-integer degree. I show that the scaling behaviour in both the orbital and parameter spaces is governed by the same sequence of eigenvalues of the linearized renormalization operator. These eigenvalues are smooth functions of the degree of the maximum of the map.

## 1   Introduction

In his seminal paper [1], Feigenbaum showed that period-doubling sequences in families of maps of the real line of the form

$$f_\lambda(x) = \lambda - |x|^d \quad (d=2,4) \tag{1}$$

have the asymptotic behaviour

$$\lambda_{k+1} - \lambda_k \equiv \Delta\lambda_k \sim \frac{a_0}{\delta(d)^k} \tag{2}$$

$$f_{\lambda_k}^{[2^k-1]}(0) \equiv \phi_k \sim \frac{b_0}{\alpha(d)^k} \tag{3}$$

as $k \to \infty$. Here $\lambda_k$ is the smallest parameter value at which $f_\lambda$ possesses a superstable $2^k$ cycle, superscript $[n]$ indicates $n$-fold composition, and $a_0$ and $b_0$

---

are constants. ($a_0$ and $b_0$ are of course dependent on $d$, but as I am not interested in their value, I suppress the dependence from the notation). $\alpha$ and $\delta$ were claimed to be *universal*, that is, dependent on $d$ only. Additionally, Feigenbaum gave an argument ([1], section 5) that the rate of convergence of $\alpha_k \equiv \phi_k/\phi_{k+1}$ to its limit is also $\delta$. This amounts to claiming that equation (3) is the first term in an expansion

$$\phi_k \sim \frac{b_0}{\alpha(d)^k} + \frac{b_1}{(\alpha(d)\delta(d))^k} + \dots . \tag{4}$$

However, Feigenbaum had evidence was that this is true for $d = 2$ only. The present study arose out of a desire to understand this discrepancy, and more generally, the dependence on $d$ of the spectrum of scaling exponents. The case $d = 2$ has been studied by Mao and Hu [9], Liu and Young [8] and by Reick [11]. Here I present results of a numerical study of the problem, and a plausibility argument to justify the assumptions used the numerical study.

## 2 Numerical results

We postulate the following forms for the behaviour of $\Delta\lambda_k$ and $\phi_k$:

$$\Delta\lambda_k = \sum_{i=0}^{\infty} \frac{a_i}{\delta_i^k} \tag{5}$$

$$\phi_k = \sum_{i=0}^{\infty} \frac{b_i}{\alpha_i^k}. \tag{6}$$

To ascertain the validity of these expansions, I accurately computed from four to eight of the quantities $\delta_i$ and $\alpha_i$ for about 50 values of $d$ between 1 and 10. Here $\delta_i$ and $\alpha_i$ are the constants to be determined, with $\delta_0 = \delta$ and $\alpha_0 = \alpha$. The exponents are named in order of increasing magnitude. I found it necessary to compute typically 15 to 20 superstable parameter values $\lambda_k$, for which I used a Newton iteration method [3]. The use of high precision arithmetic (50 to 100 decimal places) becomes essential, since up to $2^{20}$ iterations of $f_\lambda$ are being computed, and there is otherwise an unacceptable accumulation of roundoff errors.

I then estimated the exponents by the method described by Mao and Hu [9], again using high precision techniques. I illustrate this by the case of the $\delta$ scaling exponents, the case of $\alpha$ being exactly analogous. The problem reduces to solving a linear system and a polynomial equation, as follows. Assuming that $N$ scaling exponents are desired, $2N - 1$ values of $\Delta\lambda$ are required. Let us call these $\Delta\lambda_{i+1}, \dots, \Delta\lambda_{i+2N}$. Eliminating the constants $a_i$ from the definition of the scaling

exponents results in

$$\begin{bmatrix} -\Delta\lambda_{i+1} & +\Delta\lambda_{i+2} & \cdots & (-1)^N \Delta\lambda_{i+N} \\ -\Delta\lambda_{i+2} & +\Delta\lambda_{i+3} & \cdots & (-1)^N \Delta\lambda_{i+N+1} \\ \vdots & \vdots & \vdots & \\ -\Delta\lambda_{i+N} & +\Delta\lambda_{i+N+1} & \cdots & (-1)^N \Delta\lambda_{i+2N} \end{bmatrix} \begin{bmatrix} t_1 \\ t_2 \\ \vdots \\ t_N \end{bmatrix} = \begin{bmatrix} -\Delta\lambda_i \\ -\Delta\lambda_{i+1} \\ \vdots \\ -\Delta\lambda_{i+N-1} \end{bmatrix} \quad (7)$$

where $t_1 = \sum_{i=1}^{N} \delta_i$, $t_2 = \sum_{i \neq j}^{N} \delta_i \delta_j$ etc. This equation is solved for $\{t_1, t_2, \ldots, t_N\}$, and the roots of the polynomial

$$x^N - x^{N-1} t_1 + \cdots + (-1)^N t_N \quad (8)$$

are found. The $N$ roots are the desired scaling exponents $\delta_i$. This calculation can also be viewed as the construction of a Padé approximant to the analytic function $f(x) = \sum_{i=1}^{\infty} \Delta\lambda_i x^i$, and the subsequent determination of the poles $\delta_i$ of $f(x)$. This point of view suggests that in the case of integer $d$, where $\delta$ is known to very high accuracy [4], elimination of the pole $\delta$ by multiplication of $f(x)$ by $1 - x/\delta$ might achieve greater accuracy in the calculation of the higher poles $\delta_i$. However, I have not found a sufficient increase in precision to make this procedure worthwhile.

A selection of numerical values are shown in Figure 1, and the full results are shown in Figures 2-5. For reasons that will become apparent, it is most useful to consider $\delta_i/\delta$ and $\alpha_i/\alpha, i = 1, 2, \ldots$, rather than the exponents themselves. Only as many decimal places as are believed correct are quoted.

| $d = 2$ | | |
|---|---|---|
| $\delta = 4.6692, \alpha = -2.5029$ | | |
| $\delta_i/\delta$ | $\alpha_i/\alpha$ | origin |
| 4.6692 | 4.6692 | $\delta$ |
|  | 6.2645 | $\alpha^2$ |
| -8.0872 | -8.0872 | $\mu_1^{-1}$ |
| -17.4499 | -17.4499 | $\mu_2^{-1}$ |
| 21.8014 | 21.8 | $\delta^2$ |
|  | 29.9 | $\delta\alpha^2$ |
| -37.76 |  | $\delta\mu_1^{-1}$ |
| 64 |  | $\mu_1^{-2}$ |
| -84 |  | $\delta\mu_2^{-1}$ |
| $\mu_1^{-1} = -8.087165, \mu_2^{-1} = -17.4499$ | | |

| $d = 4$ | | |
|---|---|---|
| $\delta = 7.2846, \alpha = -1.6903$ | | |
| $\delta_i/\delta$ | $\alpha_i/\alpha$ | origin |
| 3.4266 | 3.4266 | $\mu_1^{-1}$ |
| -3.9114 | -3.9114 | $\mu_2^{-1}$ |
| -6.4283 | -6.4283 | $\mu_3^{-1}$ |
| 7.2846 | 7.3 | $\delta$ |
|  | -13 | $\delta_1\delta_2$ |
| $\mu_1^{-1} = 3.4266, \mu_2^{-1} = -3.9114$ | | |
| $\mu_3^{-1} = -6.4282$ | | |

| $d = 6$ | | |
|---|---|---|
| $\delta = 9.2962, \alpha = -1.4677$ | | |
| $\delta_i/\delta$ | $\alpha_i/\alpha$ | origin |
| 2.6949 | 2.6949 | $\mu_1^{-1}$ |
| -2.9242 | -2.9242 | $\mu_2^{-1}$ |
| -4.7499 | -4.7495 | $\mu_3^{-1}$ |
| 4.8100 | 4.8112 | $\mu_4^{-1}$ |
| -6.8 | -6.8 | $\mu_5^{-1}$ |
| 7 | 7.3 | $\mu_1^{-2}$ |
| $\mu_1^{-1} = 2.6949, \mu_2^{-1} = -2.9242$ | | |
| $\mu_3^{-1} = 4.7499, \mu_4^{-1} = 4.8100$ | | |
| $\mu_5^{-1} = -6.901$ | | |

| $d = 8$ | | |
|---|---|---|
| $\delta = 10.948, \alpha = -1.3580$ | | |
| $\delta_i/\delta$ | $\alpha_i/\alpha$ | origin |
| 2.3285 | 2.3285 | $\mu_1^{-1}$ |
| -2.4620 | -2.4620 | $\mu_2^{-1}$ |
| 3.6309 | 3.63 | $\mu_3^{-1}$ |
| -3.94 | -3.9 | $\mu_4^{-1}$ |
| -4.992 | -5 | $\mu_5^{-1}$ |
| 5.463 | 5.4 | $\mu_1^{-2}$ |
| $\mu_i$ not computed | | |

Figure 1: The dominant scaling exponents for even integer $d$. Here 'origin' refers to the explanation of the numerical values in terms of the eigenvalues $\mu_i$, and the gaps are unobserved scaling exponents.

# 3 Theory

I believe that all the above results can be understood by an extension of an argument of Feigenbaum [2]. I define the usual doubling operator $T$, by

$$(Tf)(x) = \alpha f \circ f(\alpha^{-1} x), \tag{9}$$

where $\alpha$ is the (assumed known) value appropriate to the degree $d$ of the maximum of $f$. A Taylor expansion in $\lambda$ of $f_\lambda$ about $\lambda_\infty$ gives

$$f_\lambda = f_{\lambda_\infty} + (\lambda - \lambda_\infty)\frac{\partial f_\lambda}{\partial \lambda} + \cdots \tag{10}$$

$$= g_0(x) + (\lambda - \lambda_\infty)h_0(x) + \cdots. \tag{11}$$

$g_k$ and $h_k$ may now be defined by

$$(T^k f_\lambda)(x) = g_k(x) + (\lambda - \lambda_\infty)h_k(x) + \cdots, \tag{12}$$

and it is known from the work of Feigenbaum [2] that as $k \to \infty$

$$g_k(x) \to g(x) \tag{13}$$

$$h_k(x) \to (DT_g)(h_{k-1})(x). \tag{14}$$

Here $g(x)$ satisfies $Tg = g$, $DT_g$ is the linearization of $T$ about $g$, and $\alpha^{-1} = g(0)$. It is known that Feigenbaum's $\delta$ is the dominant eigenvalue of $T_g$, and governs the approach of $\lambda_k$ to $\lambda_\infty$. I will now compute the effect of the subdominant eigenvalues.

Let us expand $h_0(x)$ in eigenvectors $e_j(x)$, $(j = 1, 2, \ldots)$ of $DT_g$. Letting $DT e_j = \mu_j e_j$ with $e_j(0) = 1$ and $\mu_1 = \delta$ gives

$$h_0(x) = \sum_{j=1}^{\infty} a_j e_j(x) \tag{15}$$

$$h_k(x) = a_1 \delta^k [e_1(x) + \sum_{j=2}^{\infty} \frac{a_j}{a_1}(\frac{\mu_j}{\mu_1})^k e_j(x)]. \tag{16}$$

Thus as $k \to \infty$,

$$\alpha^k f_{\lambda_k}^{2^k}(\alpha^{-k} x) \sim g(x) + (\lambda_k - \lambda_\infty) a_1 \delta^k [e_1(x) + \sum_{j=2}^{\infty} \frac{a_j}{a_1}(\frac{\mu_j}{\mu_1})^k e_j(x)], \tag{17}$$

and so

$$\lambda_\infty - \lambda_k \sim a_1^{-1} \delta^{-k} [1 + \sum_{j=2}^{\infty} \frac{a_j}{a_1}(\frac{\mu_j}{\mu_1})^k]^{-1}. \tag{18}$$

Inverting the sum in square brackets shows that

$$\lambda_\infty - \lambda_k \sim \sum_{j=1}^{\infty} b_j \delta_j^k, \tag{19}$$

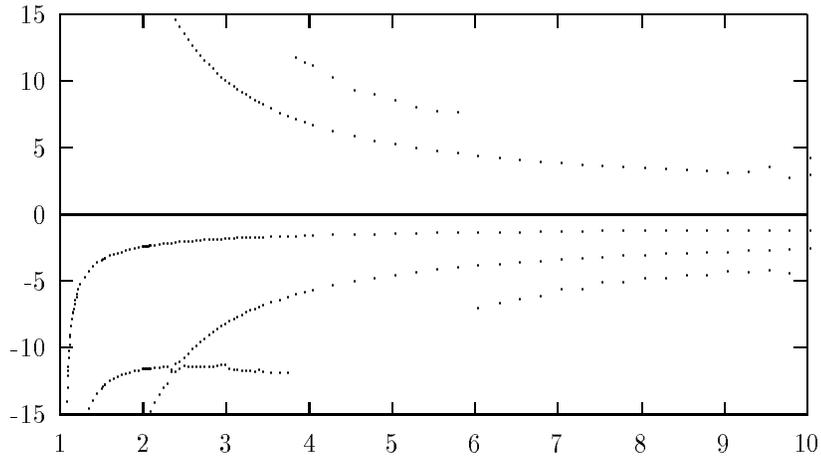

Figure 2: The scaling exponents $\alpha_i$ vs. d, $i = 0, 1, 2, 3$.

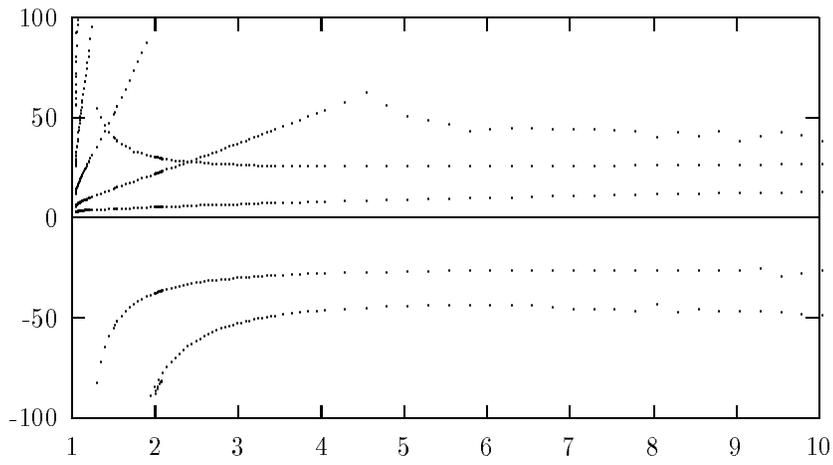

Figure 3: The scaling exponents $\delta_i$ vs. d, $i = 0, 1, 2, 3, 4$.

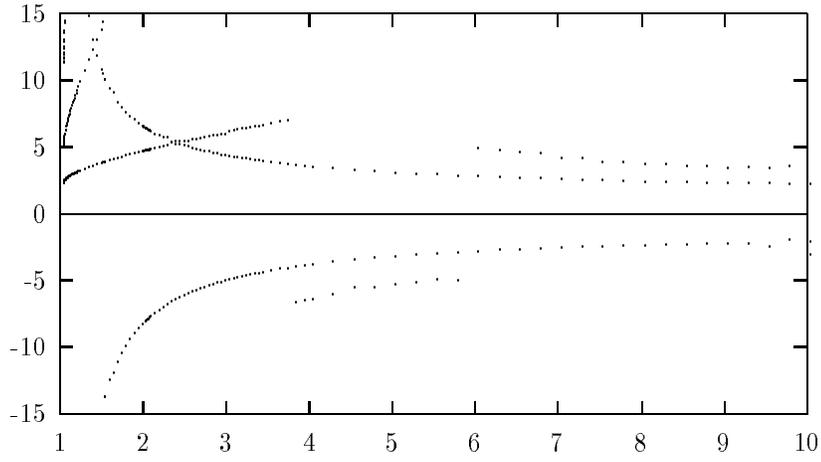

Figure 4: The ratios $\alpha_i/\alpha$ vs. $d, i = 0, 1, 2, 3$.

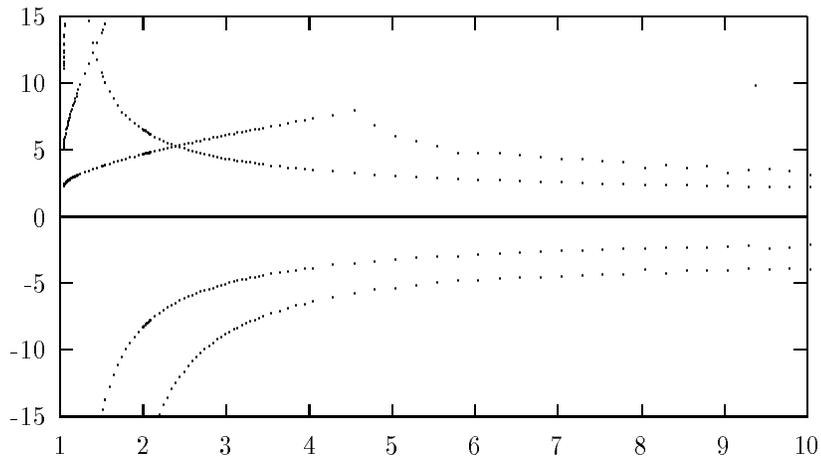

Figure 5: The ratios $\delta_i/\delta$ vs. $d, i = 0, 1, 2, 3, 4$.

where each $\delta_j$ is of the form $\delta \prod_{k=1}^{\infty} \mu_k^{-m_k}$, ($m_k$ non-negative integers), and $\{b_j\}$ are constants which are complicated functions of $\{a_j\}$.

This completes the computation of $\delta_j$ in terms of $\{\mu_j\}$. It is clear that the behaviour can be pictured as an approach of $f_{\lambda_k}$ to $f_{\lambda_\infty}$, which for an arbitrarily chosen parameterisation of $f_\lambda$, will not be along the eigendirection corresponding to $\delta$. Nevertheless, the more general approach is still governed by the largest eigenvalues of $DT_g$.

I now perform a similar calculation for $\alpha_j$. Applying the doubling operator $2^{k-1}$ times to $f_{\lambda_k}$ gives

$$\alpha^k f_{\lambda_k}^{2^{k-1}}(\alpha^{-(k-1)}x) \sim g(x) + (\lambda_k - \lambda_\infty)a_1\delta^{k-1}[e_1(x) + \sum_{j=1}^{\infty} \frac{a_j}{a_1}(\frac{\mu_j}{\mu_1})^{k-1} e_j(x)]. \quad (20)$$

Thus

$$\alpha^k \phi_k \sim 1 + (\lambda_k - \lambda_\infty)a_1^{-1}\delta^k[1 + \sum_{j=1}^{\infty} \frac{a_j}{a_1}(\frac{\mu_j}{\mu_1})^k]^{-1} \quad (21)$$

(recalling that $\phi_k \equiv f_{\lambda_k}^{2^{k-1}}(0)$). Now performing an inversion as above gives

$$\phi_k \sim \alpha^{-k}[1 + \sum_{j=1}^{\infty} c_j \delta_j^k], \quad (22)$$

where $c_j$ are constants and $\delta_j$ are the same scaling exponents as above. The important result is that the one set of eigenvalues $\mu_j$ governs both the parameter *and* the orbital scaling. The latter simply have an extra factor of $\alpha$.

There is a curious phenomenon in the case $d = 2$. The largest eigenvalue less than one takes the value exactly $\alpha^{-2}$. In other words, there is an eigenvalue crossing at $d = 2$. As expected, the effect of the eigenvalue is observed in the orbital scaling, but surprisingly, it is *not observed* in the parameter scaling. I conjecture that this is because the corresponding amplitude $a_i$ is zero (or very small). This effect nevertheless deserves more detailed study. This appears to be the only case where $\mu_j = \alpha^{-k}$ for any $j$ and $k$, though there are close approaches at $d = 4, 6$ and $8$. These close exponents make the numerical calculation difficult, and this explains the oscillations seen in the graphs at larger $d$. I believe that a more accurate calculation would show that all the exponents are monotonic functions of $d$.

The eigenvalues $\mu_j$ can be computed independently for integer $d$ via a numerical approximation to $DT_g$ [4]. This can be achieved this by first computing a finite-dimensional approximation to $g(x)$ in the form $g(x) = \sum_0^N g_i x^{di}$. This is found by solving $Tg - g = 0$ by a Newton iteration in the space of Taylor coefficients $g_i$. Similarly $DT_g$ is represented by a $N \times N$ matrix, and all its eigenvalues computed. The fact that an infinite-dimensional problem has been truncated to finite dimension means that some extra eigenvalues are introduced, but the relevant eigenvalues are usually readily identified by comparison with the bifurcation data. It is known that $\alpha^{-k}$ is an eigenvalue for integer $k$ [1], and Feigenbaum conjectured that these are all the eigenvalues of modulus $< 1$. It is clear now that this is not the case. The eigenfunctions corresponding to even integer $k$ are even, so that for odd integer $d$, our expansions cannot represent the eigenfunctions, and the eigenvalues are not found. However, this is not a difficulty as these eigenvalues are not relevant to the bifurcation data.

I list a few of the largest eigenvalues $\mu_i$ for even integer $d$ at the bottom of the tables in Figure 1. Observe that eigenvalues *not* of the form $\alpha^{-k}$ are also present. This then is the explanation of the the discrepancy between the behaviour at $d = 2$ and $d = 4$ mentioned in the introduction.

Of interest is the question of the limits $d \to 1$ and $d \to \infty$ of the scaling spectrum. Collet *et al* [6] have shown that $\delta \to 2$ as $d \to 1$. Eckmann and Wittwer [7] have shown that $\delta \to \approx 30$ as $\delta \to \infty$. On the question of this limit, see also [12]. It is also known [10] that $\alpha \to -\infty$ as $d \to 1$, and that $\alpha \to -1$ as $d \to \infty$. Unfortunately the perturbation method (about $d = 1$) used by Collet *et al* gives only the behaviour of the largest eigenvalue $\delta$. It will be observed from the graphs that all the eigenvalues appear to be smooth functions of $d$, despite the fact that there are crossings in the exponents. Notice also that powers of $\delta$ appear in the scaling spectrum, and that these dominate the behaviour for $d$ close to 1. (These powers of $\delta$ can be seen as the rapidly increasing functions near $d = 0$ in Figure 3.) Conversely, the behaviour for large $d$ is governed by a number of eigenvalues which approach each other in magnitude. In fact, these eigenvalues appear to occur in pairs of approximately equal magnitude but opposite sign. The numerical calculations suggest that the following limits exist, and take the very approximate numerical values given:

$$\lim_{d \to \infty} \delta_1/\delta \approx 2 \qquad (23)$$

$$\lim_{d \to \infty} \delta_2/\delta \approx -2 \qquad (24)$$

$$\lim_{d \to \infty} \alpha_1/\alpha \approx 2 \qquad (25)$$

$$\lim_{d \to \infty} \alpha_2/\alpha \approx -2 \qquad (26)$$

# 4  Conclusion

I have shown that all observable aspects of the period-doubling sequence in families of maps of the form $\lambda - |x|^d$ are governed by the eigenvalues of the linearized period-doubling operator. This generalizes the classical Feigenbaum scaling. The same is no doubt true in the case of complex maps $\lambda - z^m$ ($m$ integer) in the case of $n$-tupling ($n = 2, 3, 4, \ldots$). I have performed some preliminary calculations for the cases of period- doubling and period-tripling and $m = 2, 3, \ldots, 8$ which confirm this.

# 5  Acknowledgements

I am indebted to Reinout Quispel for suggesting this problem, and to Colin Thompson, Oreste Piro and Thomas Prellberg for productive discussions. I am also grateful to John Roberts for comments on an early draft of this paper. The multiple precision computations were performed with the Fortran package MP-FUN, written by David Bailey [13].

`sun:tex/correct.tex`